\documentclass[12pt,preprint]{emulateapj}  
\usepackage{graphicx,natbib}

\begin{document}

\title{The Properties of X-ray Cold Fronts in a Statistical Sample of Simulated Galaxy Clusters}
\author{Eric J. Hallman\altaffilmark{1,2,3}, Samuel W. Skillman\altaffilmark{2,4}, Tesla E. Jeltema\altaffilmark{5}, Britton D
. Smith\altaffilmark{2}, Brian W. O'Shea\altaffilmark{6}, Jack O. Burns\altaffilmark{2}, Michael L. Norman\altaffilmark{7}}

\altaffiltext{1}{National Science Foundation Astronomy and
  Astrophysics Postdoctoral Fellow}

\altaffiltext{2}{Center for Astrophysics and Space Astronomy,
  Department of Astrophysics and Planetary Sciences, 
  University of Colorado at Boulder, Boulder, CO 80309}

\altaffiltext{3}{Institute for Theory and Computation,
  Harvard-Smithsonian Center for Astrophysics, Cambridge, MA 02138; ehallman@cfa.harvard.edu}
\altaffiltext{4}{Department of Energy Computational Science Graduate Fellow}
\altaffiltext{5}{Morrison Fellow, UCO/Lick Observatories, 1156 High St., Santa Cruz, CA 95064}
\altaffiltext{6}{Department of Physics \& Astronomy and Lyman Briggs College, Michigan State University, East Lansing, MI, 488
24}

\altaffiltext{7}{Center for Astrophysics and Space Sciences,
University of California at San Diego, La Jolla, CA 
92093}

\begin{abstract}
We examine the incidence of cold fronts in a large sample of galaxy
clusters extracted from a (512$h^{-1}$Mpc) hydrodynamic/N-body
cosmological simulation with adiabatic gas physics computed with the Enzo adaptive mesh
refinement code. This simulation contains a sample of roughly 4000
galaxy clusters with M$\geq 10^{14} M_{\odot}$ at $z$=0. For each simulated
galaxy cluster, we have created mock 0.3-8.0 keV X-ray observations and
spectroscopic-like temperature maps. We have searched these maps with
a new automated algorithm to identify the presence of cold fronts
in projection. Using a threshold of a
minimum of 10 cold
front pixels in our images, corresponding to a total comoving length
$L_{cf} > 156h^{-1} kpc$, we find that roughly 10-12\% of all projections
in a mass-limited sample would be classified as cold front clusters. Interestingly, the fraction of clusters
with extended cold front features in our synthetic maps of a
mass-limited sample trends only weakly with
redshift out to $z$=1.0. However, when using different selection
functions, including a simulated flux limit, the trending with redshift changes
significantly.  The likelihood of finding cold
fronts in the simulated clusters in our sample is a strong
function of cluster mass. In clusters with  $M>7.5
\times 10^{14} M_{\odot}$  the cold front fraction is
40-50\%. We also show that the presence of cold
fronts is strongly correlated with disturbed morphology as measured by
quantitative structure measures. Finally, we find that the incidence of cold fronts in
the simulated cluster images is strongly dependent on baryonic
physics. 
\end{abstract}


\section{Introduction}\label{sec:Intro}
With the advent of high resolution X-ray imaging using space-based
observatories, in particular \textit{Chandra}, unexpected discontinuous X-ray features in galaxy clusters were
discovered. The expected appearance of shocks was supplemented by the
appearance of cold fronts, features that do not exhibit a pressure
jump, in contrast to shocks, but have jumps in surface brightness and
temperature in opposing directions \citep[for review
see][]{cf_review}.  Temperature jump ratios (from one side of the
feature to the other) in typical cold fronts range from factors of
50\% to factors of a few, with corresponding (but reversed) inferred
density jumps.  Because of the pressure continuity across these
features, they are typically described as contact discontinuities.
Early classic cold front examples
include Abell 3667 \citep{a3667} and Abell 2142 \citep{a2142}. The
properties of cold fronts have turned out to provide excellent
constraints on the physics of the intracluster medium (ICM). For
example, the extremely sharp edge (typically narrower than the Coulomb
mean free path) between the cold and hot sides of
the features places limits on the effectiveness of conduction across
the edge \citep[e.g.,][]{ettori,asai04}.  The apparent lack of fluid instability growth at the
interface also suggests some physical process that stabilizes the
edge, perhaps the draping of magnetic fields
\citep{mag3667,lyutikov,asai,taki}. 

Following the early discoveries of cold fronts, many other clusters
have been identified as hosting cold fronts
\citep[e.g.,][]{mazzotta01,1E06,dupke03,a168,johnson}. Statistics of cold fronts in clusters
have been calculated using data from the \textit{Chandra} archive
\citep{owers} and from flux-limited samples using XMM-Newton
\citep{ghizzardi06,ghizzardi} and \textit{Chandra}
\citep{twothirds}. The main results are that in flux-limited samples,
anywhere between 40\% and 87\% of clusters are found to host cold
fronts \citep{ghizzardi06, ghizzardi} depending on the redshift range explored. Also, a large fraction of cool
core clusters ($\sim$67\%) host cold fronts \citep{twothirds}.  Recent
work also suggests that the appearance of cold fronts is strongly
correlated with cluster mergers \citep{owers,ghizzardi}. 

Numerical hydrodynamic simulations of idealized cluster mergers
\citep[e.g.,][]{roett3667,heinz,takizawa,ascasibar,masburk,springfarr,zuhone} as well as
cosmological simulations \citep[e.g.,][]{nagai03,bial02,mathis,titthenr} appear
to reproduce the properties of all types of observed cold fronts
naturally in the process of mergers. Recent studies have begun to
separately classify two types of cold fronts, the first being
``merger'' cold fronts, where the system is obviously disturbed in
X-ray morphology, and clearly undergoing a current major merger, and
so-called ``sloshing'' cold fronts \citep{sloshing} in apparently
otherwise relaxed cool-core clusters. This distinction is somewhat artificial,
as simulations show that both cold front types are associated with
mergers, or in the case of sloshing, at least a close pass of a
subcluster. 

Simulation studies indicate that cold fronts arise from one of a few
physical scenarios.  First, and most obvious are the so-called
merger-type cold fronts. In these scenarios, two objects merge, at
least one a massive cluster.  In the early stages, while the merging
subclusters are approaching, their relative velocity is supersonic (in
the cluster gas) and drives shocks.  In the simplest scenario, the ram
pressure associated with the relative velocity of the subcluster in
the gas of the main cluster pushes the subcluster gas out of the dark
matter potential well.  Once the gas separates from the dark matter
potential, it expands adiabatically and cools,
resulting in the cold front feature.  In this scenario, cold fronts
should always be associated with shocks, which may also be visible in
the ICM adjacent to the cold front. Depending on when in the process
of merging the cold front is observed, the relative positions of cold
gas, dark matter, and shocks can vary.  

The sloshing type cold fronts are observed typically in cool core
clusters, and appear at small cluster radius (R$<100kpc$). These
clusters otherwise appear relaxed dynamically, leading to the separate
classification from merger type cold fronts.  There is no obvious
evidence of merging in these clusters in the X-ray images,
but simulations that reproduce cold fronts of this type show that they
are also caused by dynamical interactions. In simulations, a near pass
of a subcluster pulls the cluster mass away from the
original center position, and as the subcluster moves away, the
central baryonic material rocks back and forth in the dark matter
potential. This creates a characteristic spiral shaped cold front near
the center of the cool core cluster. 

Both the simulation work referenced here and recent
theoretical study \citep[e.g.,][]{lyutikov,birnboim09,keshet09} have
made significant progress toward understanding the process by which
cold fronts form in galaxy clusters and the detailed physics that
determines their observed properties. What has not been attempted to
date is an estimate of the predicted incidence of cold fronts in
clusters from fully cosmological numerical hydro/N-body simulations
with large (N $>1000$) samples. The appearance of merger-type cold
fronts is governed by both the merger rate of cluster scale systems,
as well as the gas dynamics and more subtle effects like projection
angle. Therefore, in order to determine how many clusters we expect to
have cold fronts at a given epoch, or mass limit, or depth of
observation, a statistical treatment is required. Though idealized
merger scenarios, or small samples simulated at high resolution are
critical to understanding the range of cold front formation, they
neglect to examine the full ranger of mergers in large scale
structure. Fully cosmological simulations, like the ones performed for
this study, create realistic merger histories, including multiple
mergers. From these simulations we can gain valuable insight about the
process of cold front formation and evolution over a large statistical
sample of galaxy clusters. 

The outline of this paper is as follows. In Section \ref{sec:simsetup},
we describe the setup and analysis of the Enzo simulations. In Section
\ref{sec:Methodology} we discuss the method of cold front
identification. In Section \ref{sec:results} we describe our major
results. In Section \ref{sec:resolution} we describe briefly the effects of
numerical resolution and baryonic physics. Finally, in Section
\ref{sec:discuss}, we discuss and summarize our results.
\section{Simulation Setup and Analysis}\label{sec:simsetup}

 The main
statistical results come from the so-called Santa Fe Light Cone (SFLC)
simulation, described in \citet{sflc}, \citet{skillman} and
\citet{hallman09}. This calculation is performed with
`Enzo'\footnote{http://lca.ucsd.edu/projects/enzo}, a publicly
available adaptive mesh refinement (AMR)
cosmology code developed by Greg Bryan and colleagues \citep{bryan97,bryan99,norman99,oshea04,
2005ApJS..160....1O}.
The specifics of the Enzo code are described in detail in these papers (and references therein).

The Enzo code couples an N-body particle-mesh (PM) solver \citep{Efstathiou85, Hockney88} 
used to follow the evolution of a collisionless dark
matter component with an Eulerian AMR method for ideal gas dynamics by \citet{Berger89}, 
which allows high dynamic range in gravitational physics and hydrodynamics in an 
expanding universe.  

 This simulation is set up as follows.  We initialize
our calculation at $z=99$ assuming a cosmological model with  $\Omega_m = 0.3$, 
$\Omega_b = 0.04$, $\Omega_\Lambda = 0.7$, $h=0.7$ (in units of 100 km/s/Mpc), 
$\sigma_8 = 0.9$, and using an \citet{eishu99} power spectrum
with a spectral index of $n_s = 1$.  The simulation is of a volume of the 
universe 512~h$^{-1}$~Mpc (comoving) on a side with a $512^3$ root grid.  The dark matter
particle mass is $7.3 \times 10^{10}$~h$^{-1}$~M$_\odot$.  The simulation was then evolved to $z=0$ with
a maximum of $7$ levels of adaptive mesh refinement (a maximum spatial resolution of 
$7.8$~h$^{-1}$ comoving kpc), refining on dark matter 
and baryon overdensities of $8.0$.  The equations of hydrodynamics were
solved with the Piecewise Parabolic Method (PPM) using the dual energy
formalism. This simulation is performed using adiabatic physics
only. It has become clear from observations that additional
non-gravitational physics is important in the evolution of the ICM.
However, this run models a large volume with high peak resolution,
which is only now becoming possible to calculate while including
additional baryonic physics.  It has the advantage of generating a
cluster sample with thousands of massive clusters, allowing us to
explore a wide range of cluster interactions that result in cold
fronts. 

\begin{figure}
\begin{center}
\epsscale{1.3}
\plotone{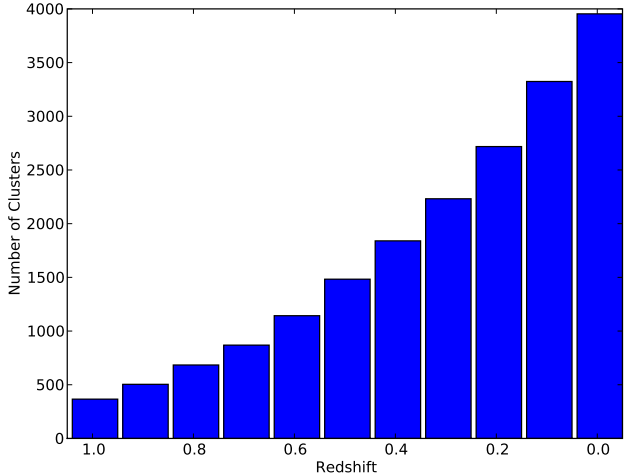}
\end{center}
\caption{Histogram of number of clusters with $M > 10^{14} M_{\odot}$
  in the comoving volume as a function of redshift. Shows the sample
  size used for the cold fronts analysis.}
\label{nclust}
\end{figure}

Analysis was performed on 11 data outputs between $z=1$ and $z=0.0$
(spaced with $\delta$z = 0.1) in an
identical way.  The HOP halo-finding algorithm \citep{eishut98} was applied to the 
dark matter particle distribution to produce a dark matter halo
catalog \citep[as in ][]{sflc}. For each output, we calculate bulk
properties (e.g., total mass, X-ray luminosity), radial
profiles, and projected images (along 3 orthogonal axes) of each simulated cluster in a variety
of physical and observable properties.
Spherically-averaged, mass-weighted radial profiles of various baryonic and dark matter
quantities including density, temperature, and pressure were generated 
for every halo in the catalog with an
estimated halo mass greater than $4 \times 10^{13}$~M$_\odot$.  These radial 
profiles were used to calculate $M_{200}$ and
$R_{200}$. $M_{200}$ refers to the total mass inside a radius of
$R_{200}$, the radius at which the overdensity average inside the
sphere centered on the cluster is 200 times the critical density.  We
then generate synthetic X-ray images for clusters in the sample with
$M_{200} >10^{14} M_{\odot}$. We then run our automated cold front finder on three
orthogonal projections of each cluster.  A
histogram of the size of the mass-limited cluster sample as a function
of redshift is shown in Figure \ref{nclust}. 

\subsection{Testing the Effect of Baryonic Physics}
For the purpose of understanding the impact of additional baryonic
physics on the cold front incidence, we have analyzed two additional
simulations. 
Each of the two simulations have identical simulation volumes, initial
conditions, and cosmology. The simulated volume is a $128h^{-1}$
comoving Mpc cube, with 256$^3$ root grid zones and 256$^3$ dark
matter particles. Each run is allowed to refine to a maximum of 5 AMR
levels, and the refinement criteria are overdensity of 8.0 in either
the dark matter or the gas (with respect to the mean on the parent
level). This gives the run a peak spatial resolution of
15.6$h^{-1}$kpc. The dark matter
particle mass is $9.0 \times 10^{9}$~h$^{-1}$~M$_\odot$ and the mean baryon mass resolution
is $1.4 \times 10^{9}$~h$^{-1}$~M$_\odot$, a factor of 8 improvement
over the SFLC run used for our main analysis.  For
these runs, we use a cosmological model with  $\Omega_m = 0.27$, 
$\Omega_b = 0.044$, $\Omega_\Lambda = 0.73$, $h=0.7$ (in units of 100 km/s/Mpc), 
$\sigma_8 = 0.9$, and using a power spectrum
with a spectral index of $n_s = 0.97$. These values are changes from our
previous run, to match WMAP5 cosmological parameters.  Therefore,
there are multiple changes to these simulations from the SFLC,
including the mass and spatial resolution, baryonic physics, and
cosmological parameters.  The most relevant changes to the cosmology
are a slight reduction in the value of $\Omega_m$ and a slight
increase in the value of $\Omega_\Lambda$. However, in this work, we endeavor to merely
compare our two runs where everything is identical \textit{except}
baryonic physics, to examine the impact. 

The two additional simulations presented here are identical in
terms of the initial conditions and simulation properties, except that one has only adiabatic and gravitational physics,
and the other includes a prescription for radiative cooling of the gas
using non-equilibrium cooling and chemistry for H and He, Cloudy
\citep{ferland} for the metal cooling, and
star formation including thermal and metal feedback from supernovae as
detailed in \citet{cen}. The cooling plus star formation and feeback
run includes a spatially uniform but time-varying
ultraviolet (UV) radiation background \citep{haardt}. We turn on the UV background
at a redshift z = 7. For the metal line cooling, we interpo-
late from a grid of data made with Cloudy. For redshifts z $>$ 7, we use a similar grid of heating and cooling
data, without the influence of the UV radiation background, assuming
collisional ionization only \citep[for more details, see][]{smith08,smith10}. Additionally, we have highly
time-resolved data outputs for these calculations, allowing us
to examine the time history and origin of cold fronts in the simulated
clusters in detail. In this paper, we examine only the
general incidence of cold fronts in these simulations, and leave a
study of the detailed properties of the cold fronts, as well as a full
convergence study, to future work.

For our simulation with purely adiabatic physics (CC-Adia), and the
identical run with additional baryonic physics (CC-OTH),
the data outputs and analysis are identical.  We perform the same
analyses as done for the SFLC simulation. 
\section{Cold Front Identification and Analysis}\label{sec:Methodology}
The identification of X-ray cold fronts in the synthetically observed
numerical clusters is done via
a relatively simple algorithm, which is designed to roughly match the
observational (by eye) identification, but in an automated way. 

For this study, we focus on the identification of X-ray cold fronts,
characterized as features which are both colder and brighter than
the surrounding intracluster medium.  This identification is made from
projected images of the clusters of the spectroscopic-like temperature ($T_{sl}$)
\citep[see][]{mazzotta04,rasia}, and the 0.3-8.0 keV X-ray surface brightness
calculated using the Cloudy code \citep{ferland}. The
spectroscopic-like temperature has been determined to be an
accurate proxy for the measured X-ray spectral temperature \citep{rasia}. We are not using the volume integral of the
spectroscopic-like temperature as in \citet{rasia}, but are using the
same weighting for the line of sight integral in order to make mock
temperature maps. This weighting has been shown to reproduce the
fitted spectral temperature better than standard emission weighting or mass
weighting. The calculation of $T_{sl}$ for our projected maps
is
\begin{equation}
T_{sl} = \frac{\int n^2T^a/T^{0.5}dl}{\int n^2T^a/T^{1.5}dl},
\end{equation}
where a=0.75. 

For each image, we
step in each row or column of the pixel map, and look for jumps in the
surface brightness and temperature which are above a threshold value,
and are opposite in sign. The jump thresholds are 
\begin{equation}
S_{x,1}/S_{x,2} > 2.0,
\end{equation}
\begin{equation}
T_{sl,2}/T_{sl,1} > 1.4,
\end{equation}
where the numerical subscripts indicate the 2 opposing sides of the
edge in temperature and surface brightness.  Note that the jump in
surface brightness is in the opposite direction of the jump in
temperature.  Observed cold fronts have temperature jumps from factors
of 50\% to factors of a few, so this threshold should be sufficient
to recover the observed distribution.  The threshold jump strengths are designed to be
roughly continuous in pressure, as the fixed band X-ray surface
brightness is only weakly dependent on temperature and depends on the
electron density squared. Therefore a surface brightness jump of 2.0
is roughly equivalent to a projected density jump of 1.4. We also
limit the identification of cold fronts to the area within a radius of
$r_{500}$ projected on the sky, where $r_{500}$ is the radius within
which the mean overdensity of the cluster is 500 times the value of
$\rho_c$, the critical density of the universe. This radius is chosen
since it is representative of the part of massive galaxy clusters that can typically be observed
with standard exposures using current X-ray telescopes. 

We expect the number of cold fronts and the size of the jumps to be a
lower limit to the true number in the simulation.  This results from a
number of factors.  First, the effects of projection on the detection
of cold fronts must be considered. Like shocks in galaxy clusters, the
detection is strongly orientation dependent. Edge-on orientations lead
to much more detectable shocks and cold fronts, face-on oriented features
are basically undetectable via these methods. Second, because of both
orientation and potential spreading of the discontinuity over several
numerical grid cells, the jump may span 2-3 pixels. Our method in that
case may miss a more gradual jump, or detect it with a smaller ratio
of temperature and surface brightness across the feature.  The first
effect (projection) is not considered directly in this study, since we
only wish to compare the observed incidence of cold fronts to the
simulated incidence, and real cluster observations suffer the same
effect. We partially compensate for this effect by using multiple
orthogonal projections of each cluster, but this doesn't impact the
fractional incidence, as we count each projection as an independent
image. The second set of effects is more challenging to correct. In
this study we have experimented with checking for features matching
our criteria across both 2 and 3 pixels.  The main result of that
experiment is that the overall statistics are not strongly dependent
on the number of pixels across which the jump is considered.  This
results from the fact that jumps that are identified in a single pixel
jump are also typically identified in 2-3 pixels as well. For this
same reason it is not effective to include all jumps of 1, 2, and 3 pixels
for instance in our analysis due to redundant detections of the same
features. For the purposes of this study, we consider the incidence of
cold fronts detected via this method to be lower limits, and the jumps
may also be underestimated (individually). Based on our tests, we expect the overall
statistics to be largely unaffected, though it will be explored in
more detail in our later work. 

Because cold fronts are extended features in the X-ray, we anticipate
that their incidence will not be strongly
dependent on grid resolution. The incidence of cold fronts may
however be somewhat dependent on the baryonic physics of the
simulation, as well as the mass resolution.  Mass resolution of the
simulation results in a higher number of resolved subclusters,
increasing the rate of mergers. Although previous work has shown that cold fronts
certainly appear in adiabatic simulations, purely adiabatic processes
may not be the only mechanisms for cold front creation. The so-called
``sloshing'' cold fronts may result from gas which has radiatively
cooled in the centers of clusters oscillating in the dark matter
potential in response to mergers. These sloshing features appear in
observed clusters with strong central entropy gradients
\citep[e.g.,][]{twothirds,ghizzardi}. In this work we make no
distinction between sloshing and merger-type cold fronts, we merely
identify all features that match the cold front criteria specified
above. In our adiabatic simulations, because we do not get strong
core entropy gradients (as are seen in the presence of strong radiative
cooling), we expect our cold fronts to be of the merger type in this simulation.
\begin{figure}
\begin{center}
\epsscale{0.8}
\plotone{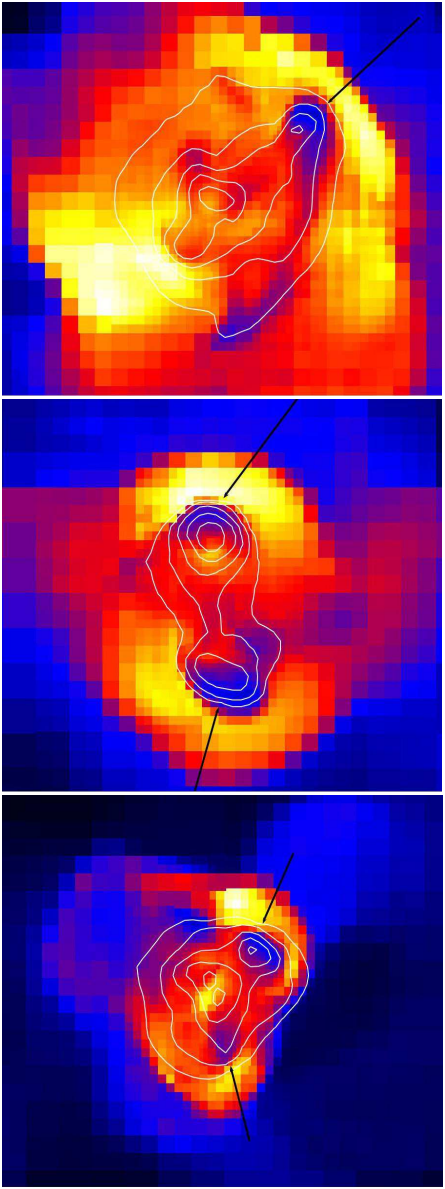}
\end{center}
\caption{Representative cold fronts from the numerically
  simulated clusters at $z$=0, with mass (from top to bottom) of 3.1$\times
  10^{15} M_{\odot}$, 1.4$\times 10^{15} M_{\odot}$, and 1.2$\times
  10^{15} M_{\odot}$. Image quantity is $T_{sl}$, contours are
  0.3-8.0 keV X-ray surface brightness. Field of view is
  roughly 3h$^{-1}$ Mpc. Image scale goes from $T\sim$3keV (blue) to
  $T\sim$10-12keV (yellow-white). Arrows indicate locations identified by the
  automated finder as cold fronts.}
\label{3panel}
\end{figure}
\epsscale{1.3}

\section{Results}\label{sec:results}
\subsection{Incidence of Cold Fronts in Simulated Clusters}
\begin{figure}
\begin{center}
\plotone{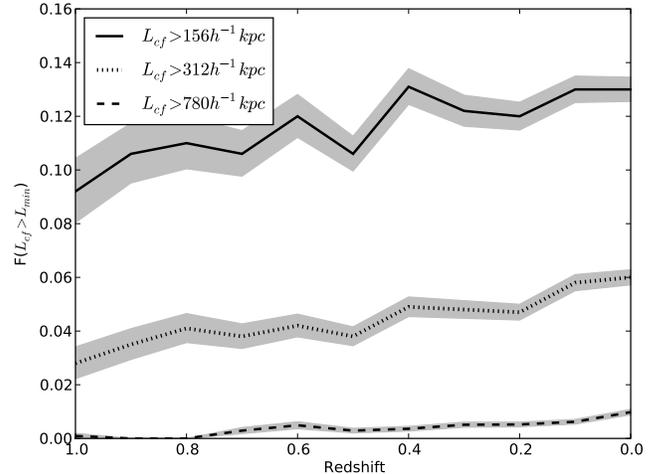}
\end{center}
\caption{Fraction of clusters displaying a cold front as a function of
  redshift. Solid line is clusters with more than 10 cold front
  pixels (total $L_{cf} >156h^{-1}$kpc comoving), dotted line is for 20 cold
  front pixels ($L_{cf} > 312h^{-1} kpc$), the dashed line represents 50
  pixels ($L_{cf} > 780h^{-1}kpc$) identified as cold
  fronts in the image.  Gray regions are 1$\sigma$ Poisson error bars. Note the
  relatively flat redshift trend. } 
\label{fraction_fz}
\end{figure}

The first result of the cold front identification is that in the
adiabatic simulation, for the sample of all clusters with $M \geq
10^{14} M_{\odot}$ out to $z=1$, 12\% of images have 10 or
more cold front pixels, 5\% have 20 or more, and roughly 1\% have 50 or more. At $z$=0 in our adiabatic simulation, we have
116 images with more than 50 cold front pixels.  In these images,
there are clear, unambiguous cold fronts spanning large areas of the
cluster projection (see Figure \ref{3panel}). Figure \ref{fraction_fz} shows the result of
counting clusters above some minimum pixel count identified as cold
fronts.  This is equivalent to measuring the comoving total length of
cold front features in each image, as each pixel has a constant
comoving size in all of these images. In the figure, we plot the trend
as a function of redshift for clusters with at least 10 pixels ($L_{cf} > 156h^{-1}$ kpc total)
identified as a cold front, 20
pixels ($L_{cf} >312h^{-1}kpc$) and 50 pixels ($L_{cf} > 780h^{-1} kpc$). These
numbers are arbitrary, and designed purely to show the general
redshift trend. It is
clear from this plot that there is only a weak trend in the incidence
of cold fronts as a function of redshift out to  $z=$1.  

The choice of a threshold in cold front pixels or comoving length of
features is fairly arbitrary, but identification of one pixel with
cold front criteria is not particularly interesting. First, it is
quite possible to have spurious cold pixels due to line-of-sight
superpositions. Additionally, cold fronts are extended
features, and so in these images at least several pixels need to be
associated in order for a true identification to take place.  

\begin{figure}
\begin{center}
\plotone{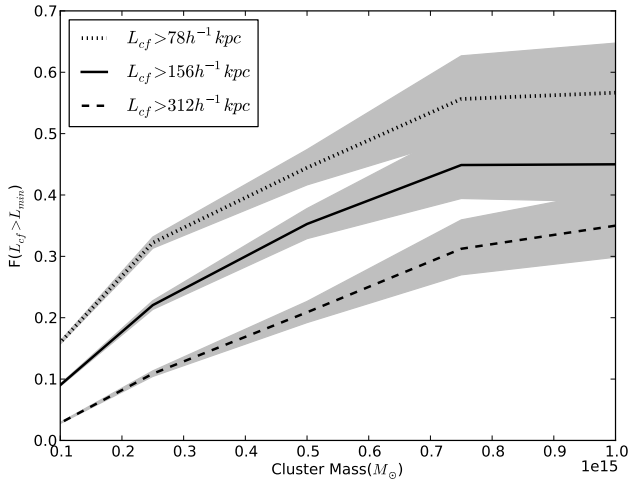}
\end{center}
\caption{Trend of cold front incidence as a function of mass for
  clusters from $1.0 \geq z \geq 0.0$.  Solid line is for clusters
  counted as cold front
  clusters if they have at least 10 identified cold front pixels,
  corresponding to a total cold front length $L_{cf} > 156h^{-1} kpc$
  comoving, within a projected radius of $r_{500}$.  These pixels are
  not necessarily contiguous. Also plotted for comparison are the
  trends with a limit of half as many pixels ($L_{cf} > 78h^{-1} kpc$, dotted
  line) and
  twice as many ($L_{cf} > 312h^{-1} kpc$, dashed line). Note the steep trend in mass, and the
  flattening at high mass ($M > 7.5 \times 10^{14} M_{\odot}$).  Gray
  regions are 1$\sigma$ Poisson error bars.}
\label{fraction_fm_all}
\end{figure}

Another interesting statistic is the incidence of cold fronts in
simulated clusters as a function of mass. For this case, we take
clusters with at least 10 identified cold front pixels, which as
described earlier is equivalent to the combined length of cold fronts
$L_{cf} >156h^{-1} kpc$. We combine the result for all clusters from $1.0
\geq z \geq 0.0$ in Figure \ref{fraction_fm_all}. What we see here is that the
likelihood of finding extended cold fronts in clusters increases with
mass.  There are several possible interpretations for this result.
The first, which is that this is a real physical trend, is that more
massive clusters form in more dense local environments, leading to
more activity from mergers and accretion, increasing the number of
cold fronts generated. A second possibility is that because smaller
clusters are smaller in projection, and we have a fixed comoving pixel
scale, they are less likely to have extended cold front features
purely because of resolution effects.  This effect is compounded by
the larger cross section for mergers for the larger clusters, generating a
higher likelihood for a merger of a given size in that projected
volume at any time.  It is legitimate to ask how the sample can have
only a weak trend in cold front incidence with redshift, yet have a strong mass
trend.  Given the hard lower mass limit at each redshift, the
distribution of clusters as a function of mass in each redshift bin is
different.  In other words, the clusters selected at each redshift
with an identical mass cutoff do not represent the same part of the
halo mass function at each redshift. This effect may contribute to the
result shown here. We expect that if cold fronts are related to merger
activity (as is indicated by simulation studies) that their incidence
\textit{should} trend with redshift, as large-scale structure grows,
and the merger rates change as a function of time.

\begin{figure}
\begin{center}
\plotone{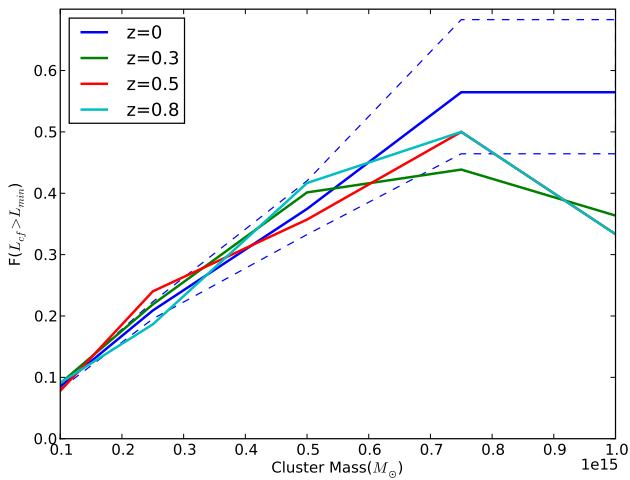}
\end{center}
\caption{Fraction of clusters with $L_{cf} > 156h^{-1} kpc$ as a
  function of cluster mass for 4 different redshift bins, $z$=0, $z$=0.3,
  $z$=0.5 and $z$=0.8.  Blue dashed lines are for 1$\sigma$ Poisson errors
  for the $z$=0 sample.  Poisson errors overlap for all redshifts at the
high mass end. }
\label{multi_z_mass}
\end{figure}

In our above calculation of the fraction of cold front clusters as a
function of cluster mass, we include clusters from all redshifts.
Does this trend with mass vary as a function of redshift?  Figure
\ref{multi_z_mass} shows the fraction of cold front clusters as a
function of mass for 4 redshift bins in the simulation. The result is
that the trend with mass is nearly identical for all redshifts, albeit
with slight shifting in the upper mass bins due to small numbers.
Effectively, clusters at all redshifts show an identical trend of
increasing fraction of clusters with cold fronts with cluster mass.

\subsection{Selection Effects}
Using a mass or flux-selected sample of galaxy clusters to study
trends as a function of redshift is an inherently flawed method, as has been pointed out by many investigators. The most well known bias
in flux-limited samples is the Malmquist bias, where your
survey progressively selects brighter objects at greater
distances. For galaxy clusters this has the effect of choosing the
higher mass objects at higher redshift.  Our sample has a bias working
in the opposite direction, which is that we are selecting a mass-limited
sample from all redshifts. Effectively we are sampling objects in
different parts of the halo mass function at each redshift, not
objects that at high redshift would be the precursors to the objects
at low redshift in our sample.  The result of such a selection is that
true evolutionary effects as a function of redshift can easily be
masked. One solution is to use a method
similar to \citet{hart}, selecting objects along an evolutionary
``road'', such that objects selected at low redshift are, in a
statistical sense, the children of the objects at higher z. 

To understand our result of the incidence of cold fronts having only
weak trending with redshift, we sort our cluster sample in two
additional ways. The first method we choose here is to use a Press-Schechter \citep{ps} mass
function analysis to select a lower mass cutoff at each redshift
corresponding to halos of the same rarity in large scale
structure. This threshold choice effectively follows the same cluster
population as a function of redshift, allowing us to see evolutionary
trends in the incidence of cold fronts. 

Our selection is done by calculating the value of $\sigma_M$, the RMS
density fluctuation, as a function of cluster mass and redshift. We
then choose a lower mass limit for our sample such that $\sigma_M$ is a
constant with redshift.  This choice is motivated by the analytic
expressions for the cluster mass function, both derived from linear
theory and fit empirically to numerical simulations. The derivations
below are taken from the cited references as well as \citet{sflc}. The comoving
number density of clusters as expressed by \citet{jenkins} is
\begin{eqnarray}
\frac{dn}{dm} (M,z) & = & -0.315 \frac{\rho_0}{M} \frac{1}{\sigma_M} \frac{d\sigma_M}{dm} \nonumber\\
& exp & \left[ -|0.61-log(D(z)\sigma_M)|^{3.8} \right]
\label{eqn-dndm}
\end{eqnarray}
where $\sigma_M$ is the RMS density fluctuation, computed on mass scale $M$ from the $z=0$ linear power spectrum \citep{eishu99}, 
$\rho_0$ is the mean matter density of the universe, defined as $\rho_0 \equiv \Omega_m \rho_c$ (with $\rho_c$ being 
the cosmological critical density, defined as $\rho_c \equiv 3 H_0^2 / 8 \pi G$), and $D(z)$ is the linear growth function, given by this fitting function:
\begin{eqnarray}
& &D(z) = \frac{1}{1+z} \frac{5 \Omega_m(z)}{2} \nonumber\\
& &\left\{ \Omega_m(z)^{4/7} - \Omega_\Lambda(z) + [1+\frac{\Omega_m(z)}{2}][1+\frac{\Omega_\Lambda(z)}{70}]\right\}^{-1}
\label{eqn-DofZ}
\end{eqnarray}
\citep{1992ARA&A..30..499C}, with $\Omega_m(z)$ and
$\Omega_\Lambda(z)$ defined in the typical way in a flat universe with
cosmological constant. For the details of the calculation, see \citet{sflc}. 

The RMS amplitude of the 
density fluctuations as a function of mass
and redshift, smoothed by a spherically symmetric window function with comoving
radius R, can be computed from the matter power spectrum using:
\begin{equation}
\sigma^2(M,z) = \int_0^\infty \frac{dk}{k} \frac{k^3}{2 \pi^2} P(k,z) |\mbox{\~{W}}_R(k)|^2
\label{eqn-sigma}
\end{equation}
where $\mbox{\~{W}}_R(k)$ is the Fourier transform of the real-space top hat smoothing function:
\begin{equation}
\mbox{\~{W}}_R(k) = \frac{3}{k^3 R^3} \left[ sin(kR)-kRcos(kR)\right]
\label{eqn-W}
\end{equation}

\begin{figure}
\begin{center}
\plotone{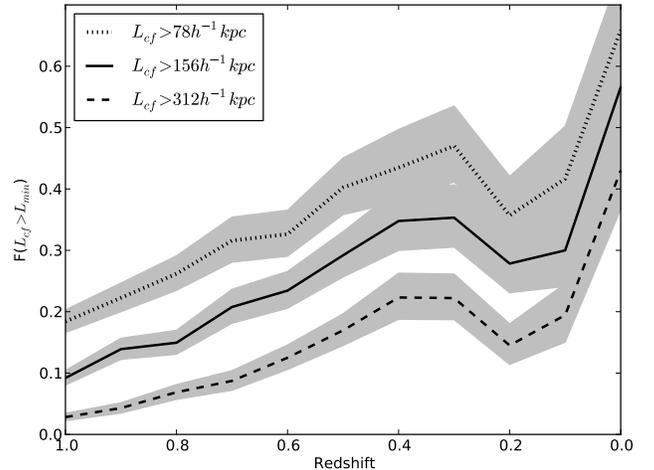}
\end{center}
\caption{Trend in redshift of clusters with the indicated $L_{cf}$ for
each line. Dotted line indicates $L_{cf} > 78h^{-1} kpc$, solid line
is for our fiducial limiting $L_{cf}> 156h^{-1} kpc$, dashed line is for $L_{cf}>312h^{-1} kpc$. This plot uses a lower mass limit in each redshift bin
drawn from a constant $\sigma_M$ limit, highlighting evolutionary
trends in the same set of clusters through time. Included are the
trends with the same limiting length of cold front pixels as in
previous Figure \ref{fraction_fm_all}. Gray
  regions indicate 1$\sigma$ Poisson error bars.}
\label{sigma_limit}
\end{figure}

We choose the value of $\sigma_M$ for our lower limit from the lower
mass limit at $z$=1, which is $M_{limit}$ = 10$^{14} M_{\odot}$.  Simply put, we use a constant $\sigma_M$ limit, and take all clusters
in the simulation above that lower mass limit at each redshift.  The
result is shown in Figure \ref{sigma_limit}.
\begin{figure}
\begin{center}
\plotone{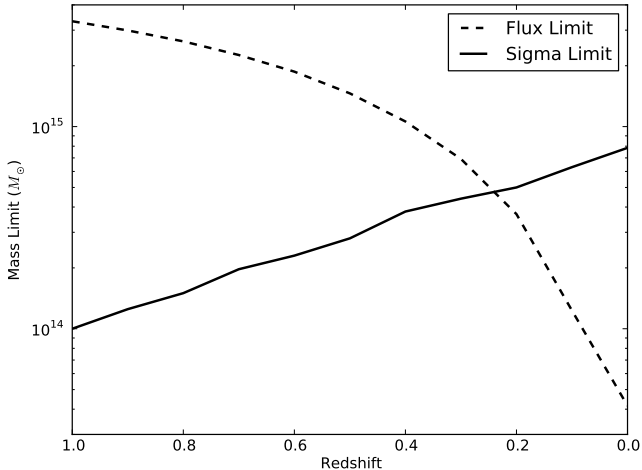}
\end{center}
\caption{Mass limits derived from the $\sigma_M$- and flux-limited
  cases described in the text. In each case, the sample of clusters
  used in the analysis are those above the specified mass limit at
  each redshift.}
\label{mass_limits}
\end{figure}

The result clearly changes the sense of the redshift trend from the
mass-limited sample. Now we see an increase in the fraction of
clusters with cold fronts as a function of time. This selection should
allow us to isolate evolutionary effects, as we are sampling the same
part of the mass function at each redshift.  The fraction generally increases as
we move forward in time, which is not particularly surprising, since
the result of our sample selection is that the lower mass limit goes
up with time.  As we have shown, the fraction of clusters with cold
fronts is a strong function of mass. What is puzzling is the
dip in the fraction at $z$=0.2. It is not entirely clear how to explain
the appearance of this feature, though it is possible that a varying
lower mass limit may create arbitrary features. 

The second additional selection mimics an observational selection by a
flux limit. Since this simulation has only adiabatic physics, and as
such does not predict the X-ray luminosity accurately, we do not use
the total X-ray emission calculated from the physical properties of
the simulated clusters.  Instead, we take a simple approach, using
observationally determined X-ray scaling relations to create a mass
limit as a function of redshift for our sample. 

We take the X-ray scaling relation fit by \citet{mantz10}, a simple
power law
\begin{equation}
l(m) = \beta_0^{lm} + \beta_1^{lm} m,
\end{equation}
where 
\begin{equation} 
l = log_{10}\left(\frac{L_{500}}{E(z) 10^{44} erg s^{-1}} \right),
\end{equation}
and 
\begin{equation}
m = log_{10}\left(\frac{E(z) M_{500}}{10^{15} M_{\odot}} \right).
\end{equation}
$E(z)$ is defined in the usual way in a flat $\Lambda$CDM
universe, 
\begin{equation}
E(z) = (\Omega_m (1+z)^3 + \Omega_{\Lambda})^{1/2},
\end{equation}
and $\beta_0^{lm}$ and $ \beta_1^{lm}$ are the fitted
parameters.
\cite{mantz10} find those parameters to be $\beta_0^{lm}$
= 0.82$\pm$0.11, and $\beta_1^{lm}$ = 1.29$\pm$0.07 by fitting to the
X-ray luminosity function. We use the flux limit from the ROSAT-ESO
Flux-Limited X-ray sample \citep[REFLEX;][]{bohringer}. In this case,
we are not interested in any particular flux limit, we simply
want to show the effect of sample selection on the cold front
statistics. The REFLEX flux limit is 2.0$\times$10$^{-12} erg s^{-1}
cm^{-2}$. At each redshift from the simulation, we calculate the X-ray
luminosity appropriate to the flux limit, then calculate the cluster
mass associated to that luminosity using the scaling relation from
\citet{mantz10}. For the $z$=0 clusters, we use $z$=0.05 to calculate the
value of $D_L$ and the flux. We compare the lower mass limits as a
function of redshift in the $\sigma_M$- and flux-limited cases in
Figure \ref{mass_limits}.
\begin{figure}
\begin{center}
\plotone{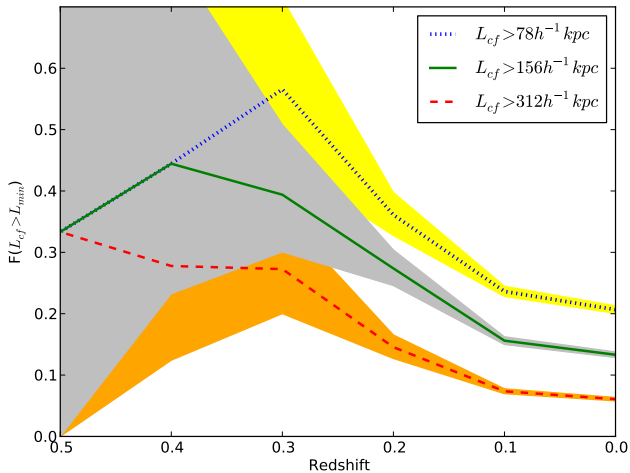}
\end{center}
\caption{Trend in redshift of clusters with the indicated $L_{cf}$ for
each line. This plot uses a lower mass limit in each redshift bin
drawn from a flux limit (from the REFLEX cluster sample) and the X-ray
scaling relations from \citet{mantz10}. Each cluster mass is converted
to a luminosity from the scaling relations, and then a flux given its
luminosity distance.  Included are the
trends with the same limiting length of cold front pixels as in
previous Figures \ref{fraction_fm_all} and \ref{sigma_limit}. This includes our fiducial
limiting $L_{cf}> 156h^{-1} kpc$ (solid green line), as well as limits of half the fiducial
length (dotted, blue line)  and twice the fiducial length (dashed, red line). Yellow, gray, and orange 
  regions are 1$\sigma$ Poisson error bars for the $L_{cf}>78h^{-1} kpc$
  line, $L_{cf}> 156h^{-1} kpc$ line and the $L_{cf}>312h^{-1} kpc$ line
  respectively.} 
\label{flux_limit}
\end{figure}

The resulting statistics of cold front incidence are
shown in Figure \ref{flux_limit}. Note that this plot only shows
fractions of clusters with cold fronts from $z$=0.5 to $z$=0, as at this
flux limit, we have no clusters in our simulation above the appropriate
mass at higher redshifts. In any case, it is clear that a flux limited
selection results in a very different picture of the incidence of cold
fronts when compared either to the mass-limited or $\sigma_M$-limited
cases.  The flux limit, of course, preferentially selects higher mass
objects at higher $z$.  As we have seen in our analysis of the incidence
of cold fronts as a function of mass, it is a strong function of mass,
so this result is not surprising.  In a flux limited sample, we should
expect the highest fraction of cold front clusters at high redshifts
(since we select the highest mass clusters there),
and it should decrease monotonically as we move to lower redshift. In
other words, given that the flux limit prefers high mass clusters at
high redshift, and we know that there are more cold fronts in high
mass clusters at all redshifts from our analysis, we expect the
highest fraction of cold front clusters at high $z$ in a flux-limited
sample. In
this selection, the total fraction of clusters with $L_{cf} >
156h^{-1}$ kpc in the full sample is $\sim$15\%, lower than typically
found in observed samples. 

\subsection{Properties of the Cold Fronts}
\begin{figure}
\begin{center}
\plotone{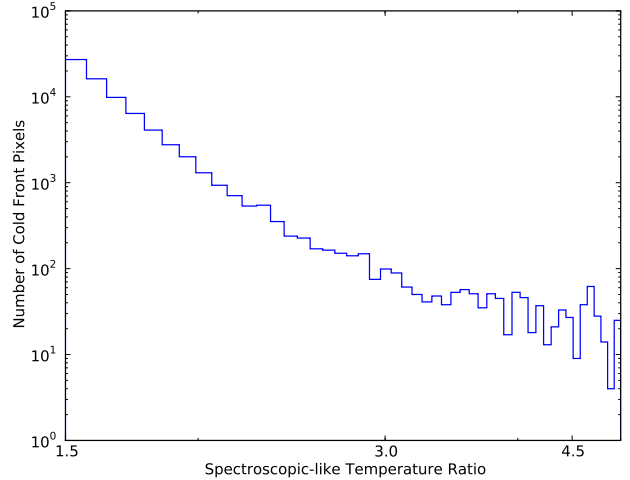}
\end{center}
\caption{Frequency of cold fronts as a function of ratio of $T_{sl}$
  ratio from one side to the other identified in all cluster images
  (inside $r_{500}$) from
  1.0$\geq z \geq $0.0.}
\label{tjump_all}
\end{figure}

\begin{figure}
\begin{center}
\plotone{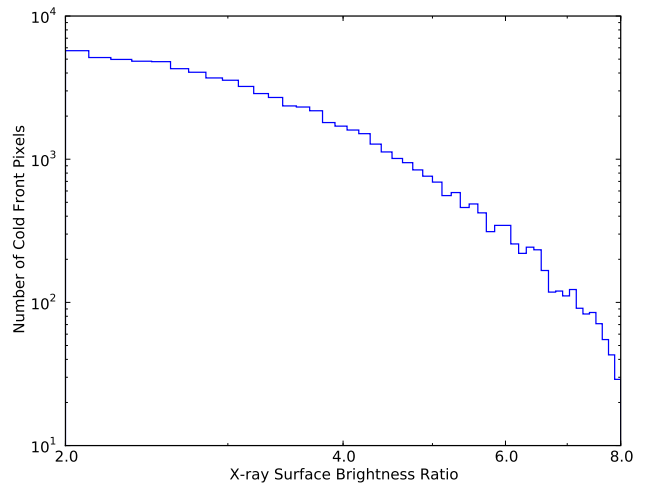}
\end{center}
\caption{Frequency of cold fronts as a function of ratio of 0.3-8.0
  keV X-ray surface brightness from one side to the other in all
  simulated cluster projections (inside $r_{500}$ in the range
  1.0$\geq z \geq $0.0.}
\label{sbjump_all}
\end{figure}

Here, we study
the properties of the identified simulated cold fronts -- in particular the jump in temperature and surface
brightness across the features. We expect from previous work looking
at the shock population \citep[e.g.,][]{ryu03,skillman}, that the
distribution of number of pixels as a function of ratio of $T_{sl}$
and 0.3-8 keV X-ray surface brightness, should be roughly a power
law.  As with shocks, in the formation of large-scale
structure, we expect more small-ratio jump features than large-ratio ones, meaning that
low velocity and temperature contrasts tend to be more plentiful than
high contrasts. With our simple cold front identifier, we indeed
find that the temperature jump frequency for all the clusters from $z$=1
to $z$=0 follow a rough power law with a break, as shown in Figure
\ref{tjump_all}. The jumps in surface brightness follow a less simple
distribution, flat at low jump ratios and rolling off at large values
as shown in Figure \ref{sbjump_all}.  A population of pressure
continuous features in the simulations should produce
a distribution of temperature and density jumps with similar
shapes. Though here we show the temperature and surface brightness
jumps, under the assumption that the soft X-ray emissivity is only
weakly dependent on temperature, the surface brightness and density
jumps have very similar shapes (though density is proportional to roughly the 
square root of X-ray emissivity).  It is unclear why the shape of the temperature
jump distribution does not qualitatively match the shape of the
surface brightness jumps. 

\begin{figure}
\begin{center}
\plotone{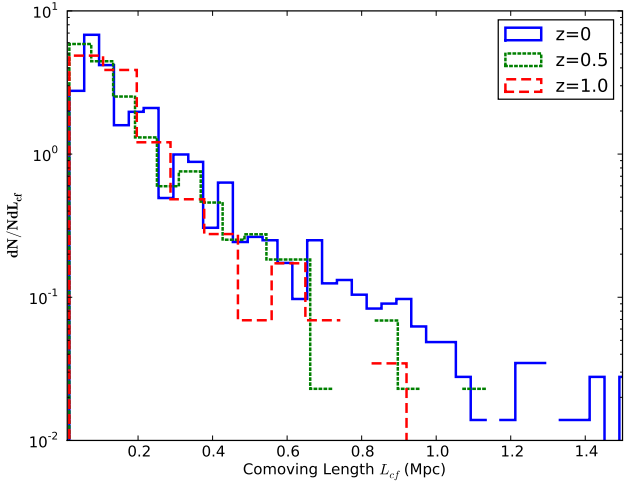}
\end{center}
\caption{Normalized probability distribution function for the comoving
length of cold fronts ($L_{cf}$) in all cluster projections at 3 redshifts as
indicated in the legend. The distribution of lengths is fairly
constant with redshift, though the high $L_{cf}$ tail lengthens at
lower redshift, due to the increased size of the largest clusters.}
\label{cf_length}
\end{figure}

Additionally, we can check the number of identified cold front pixels
per simulated cluster image. We have already described this
measurement as
equivalent to the comoving total length ($L_{cf}$) of cold front
features in the full image, though we make no effort here to determine
if the features are contiguous.  For now, this calculation gives us a
sense of the total extent of cold fronts in all the simulated cluster
images.  Figure \ref{cf_length} shows the result of this calculation,
a probability distribution function for the total comoving length of
cold fronts in the synthetic cluster maps at three
redshifts. Interestingly, the distributions at three redshifts (1.0,
0.5 and 0.0) overlap quite well.  The normalization is adjusted by the
total number of objects with cold fronts identified, but the relative
number of objects with a specific measured length of cold fronts is
more or less constant across redshift. The high $L_{cf}$ end of the
distribution grows longer at later times, presumably because the
objects progressively get larger, and therefore have larger values of
$r_{500}$, thus more area within which to find cold fronts. 

\begin{figure*}
\epsscale{1.0}
\begin{center}
\plotone{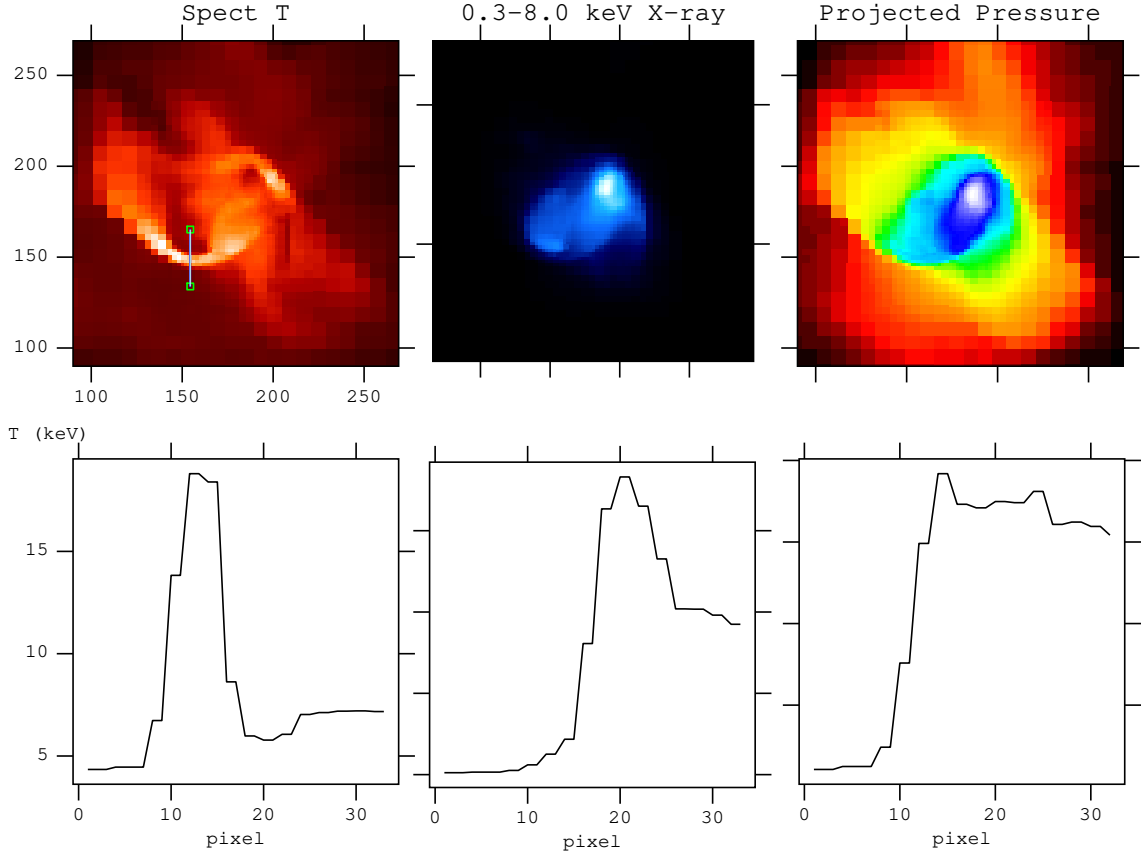}
\end{center}
\caption{Left to right: Projected images of $T_{sl}$, 0.3-8.0 keV
  X-ray surface brightness, and pressure for a cold
  front cluster with $M$ = 2$\times 10^{15} M_{\odot}$ at $z$=0 in the SFLC simulation. Also shown is the profile of
  $T_{sl}$, X-ray surface brightness and projected pressure across the shock and cold
  front in the lower panels. Location of the profile pixels is shown
  in the $T_{sl}$ image, where pixel zero is at the bottom of the
  line, and pixel 36 is at the top in the image. Pixel scale is
  15.6h$^{-1}$kpc.  Note the very hot shocked gas and the strong drop
  in temperature at the same location where the X-ray surface
  brightness goes up sharply behind the shock. Also note that at the
  shock, there is a strong pressure jump, while at the cold front
  (where the temperature drops and the surface brightness peaks) the
  pressure is roughly constant.}
\label{two_jumps}
\end{figure*}

Since cold front jumps are observed to be continuous in pressure, we check, in
projection, whether these features are consistent
with observations. If we take the projected surface
brightness jump and use it to estimate (roughly) the projected density
jump, and compare that to the corresponding projected temperature
jump, we can check to see if pressure continuity is met. The X-ray surface brightness in
the 0.3-8.0 keV band is only very weakly dependent on temperature, and
is effectively representative of the projected value of $n_e^2$. For
one of our 
simulated cold front clusters, we show the properties of a cold front
in Figure \ref{two_jumps}.  We
use the square root of the X-ray surface brightness jump as a proxy
for the projected density, and multiply it by the value of $T_{sl}$ to
form a projected pressure map. Figure \ref{two_jumps} shows the
obvious difference between the hot feature (a shock) and the cold
feature behind it (a cold front).  In the shock case, the jump in
temperature corresponds to a pressure jump, and in the cold front
case, the temperature jump is effectively continuous in pressure. Note
also the opposite direction of the jump in surface brightness and
temperature in the cold front. These jumps are typical of the shocks
and cold fronts in the simulated clusters.
\epsscale{1.3}
\subsection{Cold Fronts in Individual Clusters}
The cold fronts identified in the images of the simulated clusters
have properties consistent with those observed in real galaxy
clusters. The jumps in temperature are typically factors of 1.5-3,
with a corresponding inverse change in the X-ray surface brightness, such
that the features are continuous in pressure.  Cold fronts in the
simulated images are typically associated with shocks, as is clear in
Figure \ref{two_jumps}. While we have not specifically looked at the
observability of these features in the X-ray, it is clear from our
data that the surface brightness in the hot shocked region is
significantly lower than that in the cluster core. It is possible that
many of these shocks would not be detected in X-rays for a typical
exposure time, though the cold front would be more easily
detectable. In future work, we will explore the statistics of cold
fronts given the X-ray observability using more sophisticated
synthetic observations including instrument response and X-ray
backgrounds. 

\begin{figure}
\begin{center}
\epsscale{1.1}
\plotone{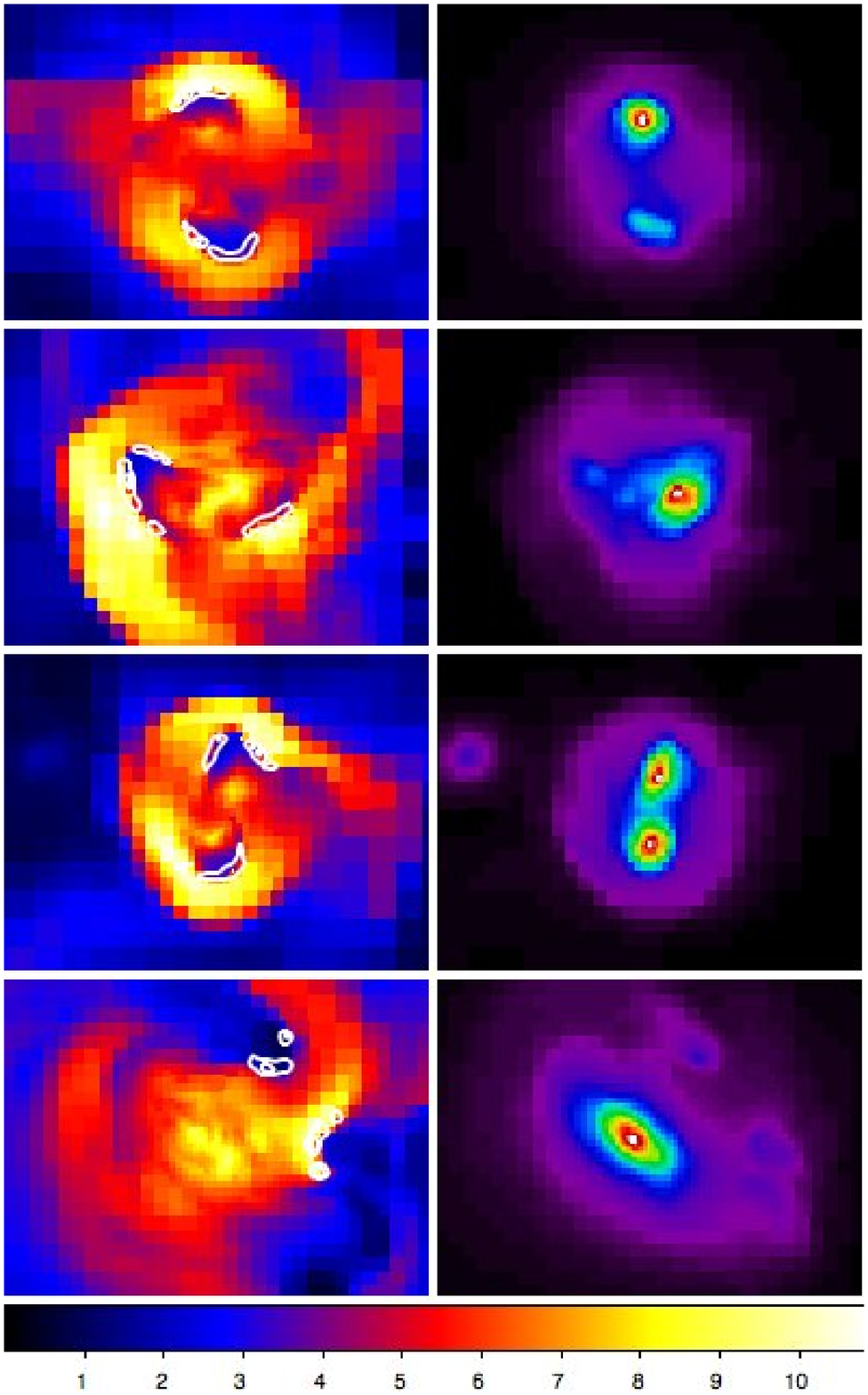}
\end{center}
\caption{Images of the simulated galaxy clusters
  with the most extended cold front structures. Left column is
   projected spectroscopic-like temperature  with cold fronts identified using the automated method described in the
  text overlaid as white contours, second column is 0.3-8.0 keV X-ray
  surface brightness. Image scale is $\sim$3 Mpc. Colorbar indicates
  temperature scale from left column in keV. Masses of the
  clusters from top to bottom are: 1.4$\times 10^{15}M_{\odot}$,
  8.5$\times 10^{14}M_{\odot}$, 4.8$\times 10^{14} M_{\odot}$,
  1.98$\times 10^{15} M_{\odot}$.}
\label{multipanel}
\end{figure}

It is also worth exploring the recent history of the simulated
clusters with the most obvious cold front features. Simulations show
that cold fronts should be a good diagnostic for recent merger activity \citep[e.g.,][]{mathis,zuhone}, both in the
case of sloshing type cold fronts and the more extended merger type
cold fronts. A multipanel image of some of the most extended cold
fronts in our simulated sample at $z$=0 are shown in Figure
\ref{multipanel}. The left column of this image shows the projected spectroscopic-like
temperature (with identified cold fronts shown as white contours), and the right column is the X-ray surface
brightness.  In these cases it is very clear that the cold front
finder has done a sufficiently accurate job of characterizing the cold
front regions. We note that these cold fronts are all of the merger
type, and not the sloshing type.  We do not see the sloshing type cold
fronts in these adiabatic clusters, and do not expect them given the
lack of radiative cooling and steep central entropy gradients. 
\epsscale{1.3}

\begin{figure}
\begin{center}
\plotone{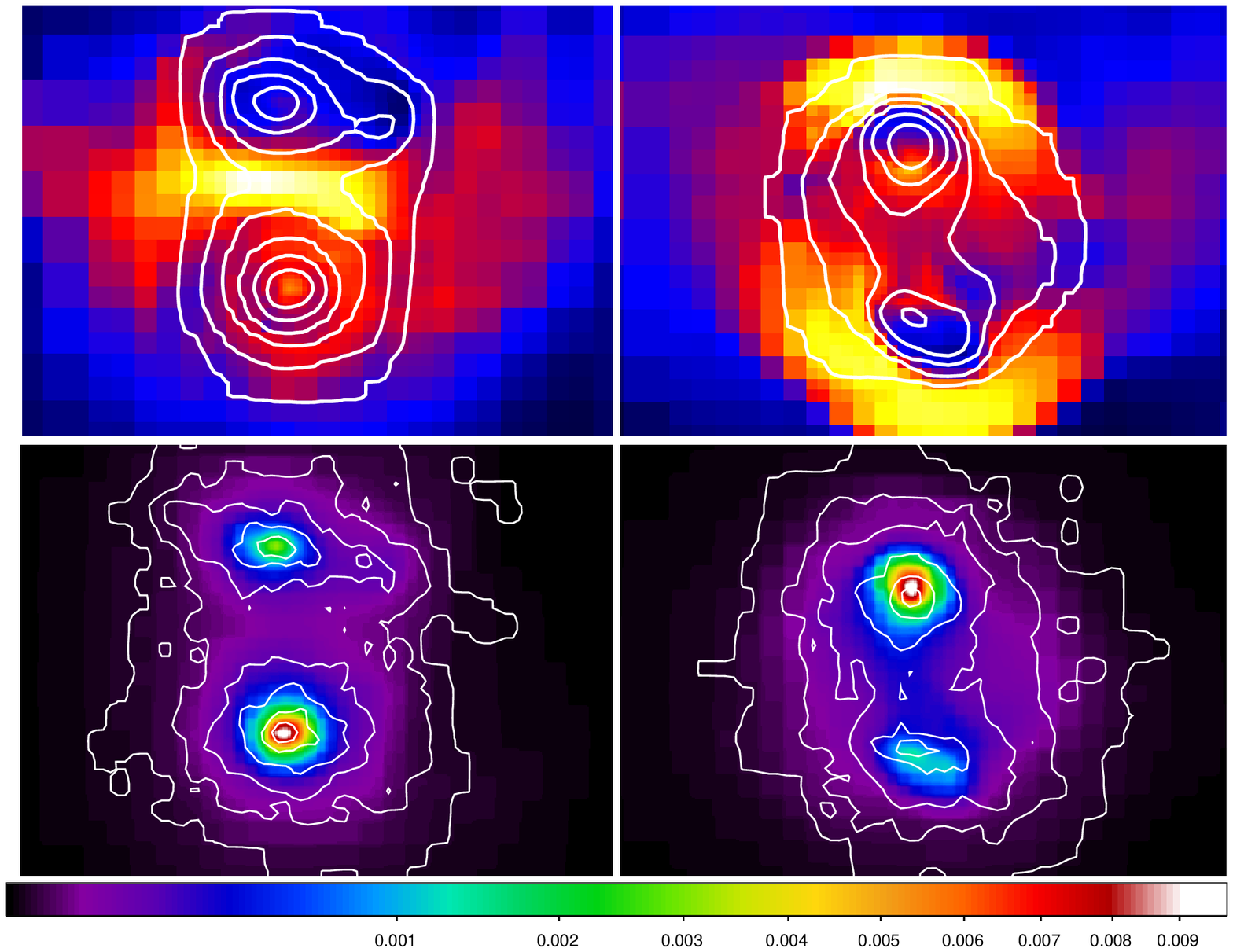}
\end{center}
\caption{Time sequence of development of a cold front in a
  1.4$\times 10^{15} M_{\odot}$ cluster (same as the cluster in the
  top panels of Figure \ref{multipanel}). Left panels are at $z$=0.1, right are at
  $z$=0. Upper panels show projected spectroscopic-like temperature in color and 0.3-8.0 keV
  X-ray surface brightness in contours. Color range in $T_{sl}$ goes
  from 3 keV (blue) to 10 keV (white).  Bottom panel is X-ray surface
brightness in color and projected dark matter density
contours. Colorbar on this image shows range of X-ray surface
brightness in $ergs/s/cm^2/pixel$ at the face of the projected simulation
box (not corrected for distance of observer) to allow a
comparison between the two images. Image
scale is 2.0 Mpc. Note the obvious merger at $z$=0.1, which creates the
cold fronts after the cores of the subclusters pass through each other.}
\label{halo33}
\end{figure}

\begin{figure*}
\begin{center}
\epsscale{1.0}
\plotone{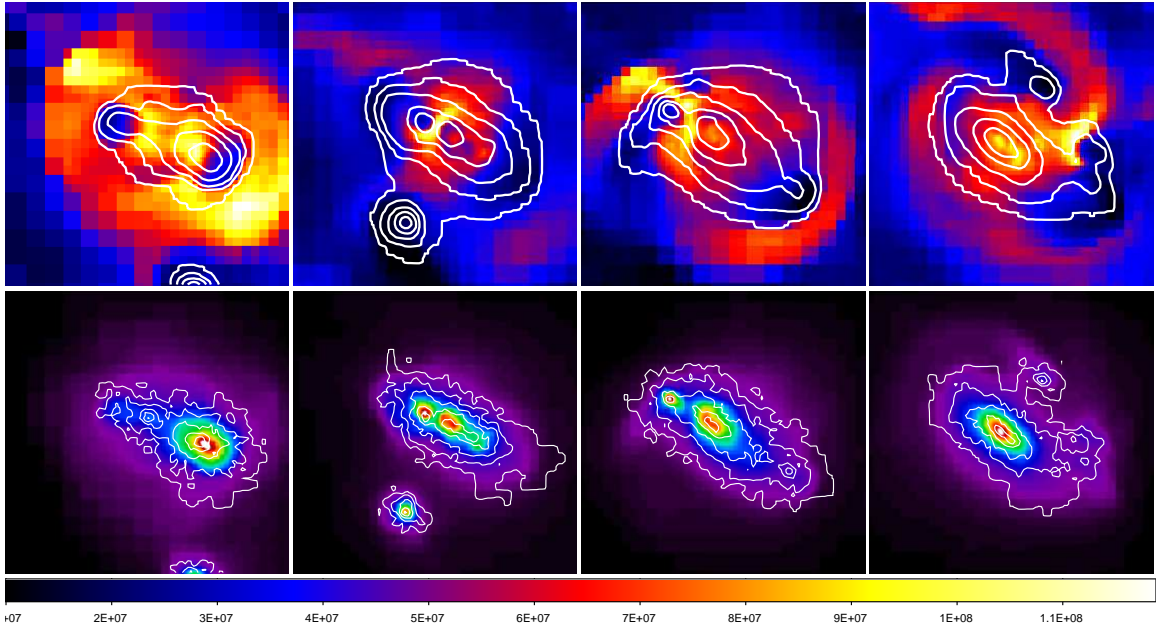}
\end{center}
\caption{Time sequence for a cold front developing in a 1.98 $\times
  10^{15} M_{\odot}$
  cluster  (same as the cluster in the
  bottom panels of Figure \ref{multipanel}). Left to right panels are forward in time, from $z$=0.3,
  $z$=0.2, $z$=0.1 and $z$=0.0. Upper and lower panels show the same
  quantities as in Figure \ref{halo33}. Image scale is 4.0 Mpc,
  colorbar is $T_{sl}$ in Kelvin for the upper panels. Note the obvious cold fronts
  in the earliest timestep, followed by a recollapse and shock, then
  an additional subgroup merges at the same time in the third
  timestep, causing the development of multiple cold fronts simultaneously.}
\label{halo6}
\end{figure*}

\begin{figure}
\begin{center}
\plotone{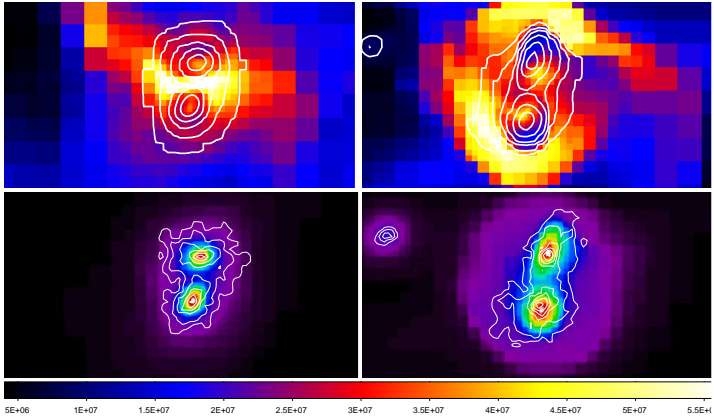}
\end{center}
\caption{Time sequence for a 4.8$\times 10^{14} M_{\odot}$ cluster (same as the cluster in the
  second from bottom panels of Figure \ref{multipanel}), right panels are
  $z$=0, left are $z$=0.1.  Image quantities are the same as Figures
  \ref{halo33} and \ref{halo6}.  Colorbar is $T_{sl}$ in Kelvin for the upper panels. Image scale is 2.0 Mpc.  Note the
  similarities to Figure \ref{halo33}, with an obvious plane of the
  sky merger in the first timestep and the classic post core-crossing
  cold fronts in the second timestep.}
\label{halo373}
\end{figure}

For three of these obvious cold front clusters, we have taken
snapshots of their recent history to illustrate the process by which
these extended cold fronts are formed in the cosmological
simulations.  Recall that for this large set of simulated clusters,
the physics in the simulation is purely adiabatic, meaning that there
is no radiative cooling.  Therefore any regions where cold and hot gas
are in close proximity result either from shocks or adiabatic
effects. The histories of these three clusters are shown in Figures
\ref{halo33},\ref{halo6} and \ref{halo373}.  In each case, we see that
the formation of the cold fronts is a result of a recent merger. 

In Figures \ref{halo33} and \ref{halo373}, we show two classic roughly
plane-of-the-sky mergers, where in the earlier timesteps, the
halos of the merger are clearly approaching each other, as
evidenced by the shock features between them. At the later epoch, we
see the subclusters after core passage, when the cold fronts have
developed. We also show the dark matter contours, and there appears to
be some mild separation of the dark matter centroids from the gas
peaks, though we have not quantified this result in this work. In
Figure \ref{halo6}, we see an array of cold fronts throughout the
history of the cluster, but in this case the dynamics are more
complicated. In the far left panels, we see a situation where there
has been an initial core passage of two subclusters. The second panels
show the approach of the two objects as they fall back together, with
an obvious shock between, while at the same time a third subcluster
approaches from the lower left. In the third panel, the initial merger
has had a second core passage, while the extra merging subcluster is now
driving a shock as it wraps around the subcluster in the upper left.
At the final stage, there are two very obvious cold fronts, but
deducing the merging history from the final snapshot could be quite
challenging.  

Cold fronts in the simulated clusters result from a
variety of dynamical scenarios, ranging from simple binary mergers to
multiple simultaneous mergers. The appearance of cold fronts and
shocks is however a generic result of all kinds of mergers in these systems.
\epsscale{1.3}

\subsection{Cold-Front Cluster Morphology}
\begin{figure}
\begin{center}
\plotone{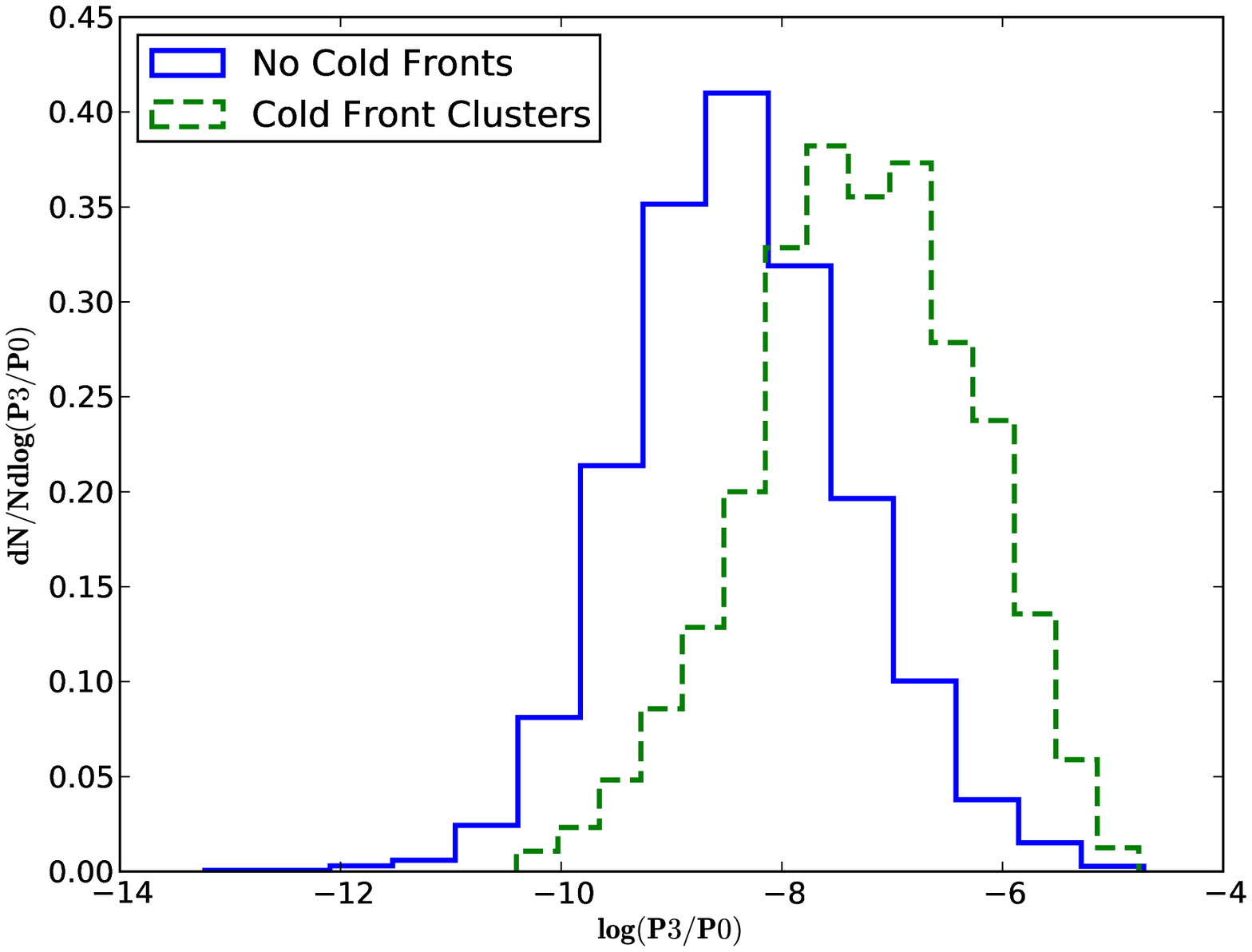}
\end{center}
\caption{Probability density function of clusters as a function of $P_3$/$P_0$ power
  ratio for each cluster image in the $z$=0 sample. Green dashed line is cold front clusters,
  blue solid line is non-cold front clusters. Cold front clusters are those
  that have a minimum comoving length of cold front pixels  $L_{cf}
  >156h^{-1}$kpc comoving. }
\label{p3}
\end{figure}
\begin{figure}
\begin{center}
\plotone{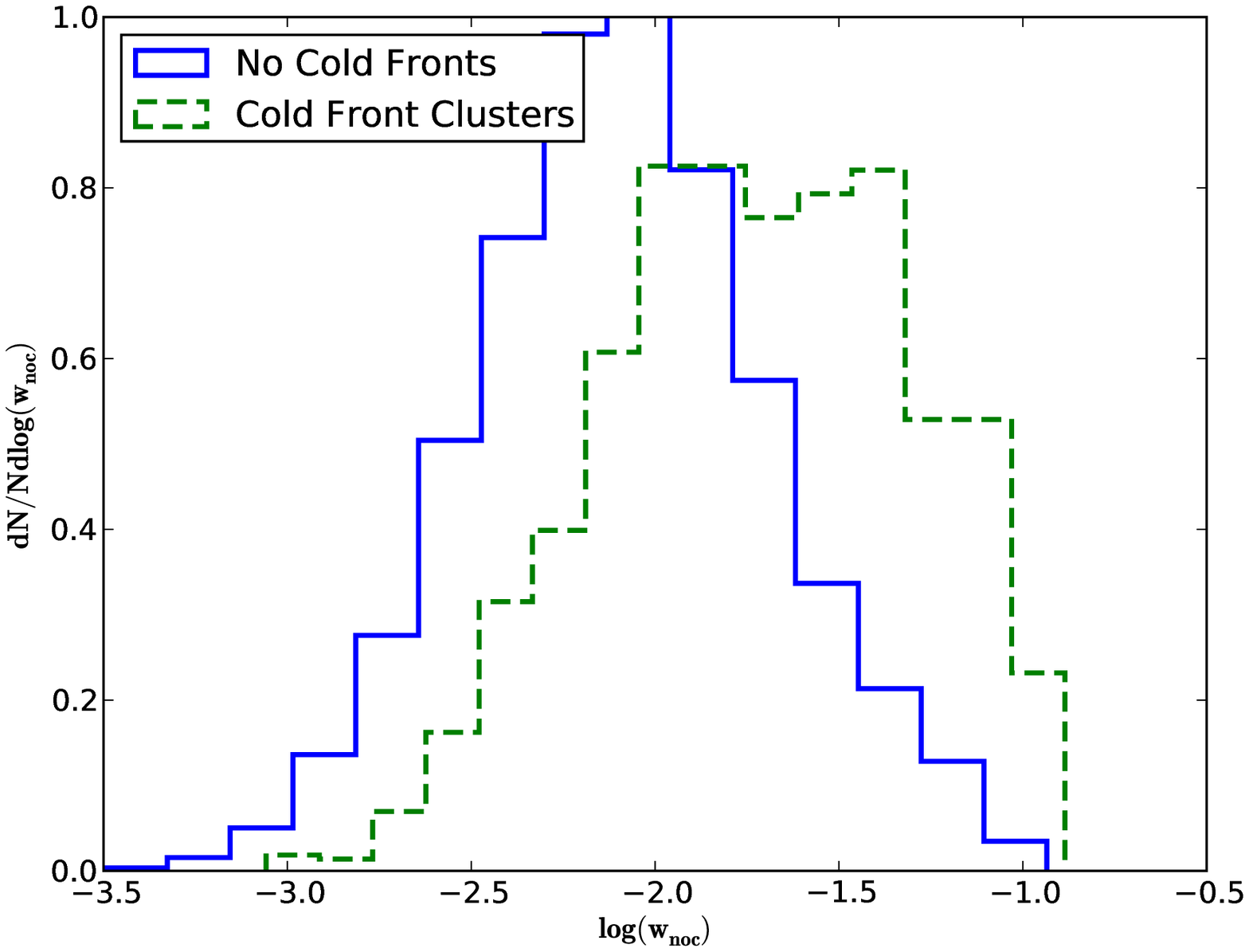}
\end{center}
\caption{Probability density function of clusters as a function of centroid shift
  for each cluster image in the $z$=0 sample. Green dashed line is cold front clusters,
  blue solid line is non-cold front clusters.  Cold front clusters are those
  that have a minimum comoving length of cold front pixels  $L_{cf}
  >156h^{-1}$kpc comoving. }
\label{p4}
\end{figure}

Recent studies of well-known cold front clusters identify a possible
relationship between the presence of X-ray cold fronts and the merger
state of the clusters \citep{owers}. There is also some indication that in
some fraction of clusters the presence of the cold front may be the
only indicator of a recent merger \citep{ghizzardi}.  To test these ideas, we have
undertaken a study of the X-ray morphology of the clusters in our
simulated samples via the measurement of power ratios and centroid
shifts of the X-ray images. These quantitative measures
are an indicator of whether the X-ray morphologies are obviously
disturbed or regular, and are a proxy for deciding whether a
cluster is dynamically active or relaxed \citep{jeltema,vent08}. 

In short,
the power ratios are the multipole moments of the X-ray surface
brightness map from a circular aperture centered on the cluster X-ray 
centroid \citep[see][]{bt95,bt96,jeltema05}. 
As described elsewhere, this method is related to the multipole
expansion of the two-dimensional gravitational potential.  This set of
equations is reproduced from \citet{jeltema}. The multipole expansion of the two-dimensional gravitational potential is
\begin{eqnarray}
\Psi(R,\phi) = -2Ga_0\ln\left({1 \over R}\right) -2G
\sum^{\infty}_{m=1} {1\over m R^m}\nonumber\\
\left(a_m\cos m\phi + b_m\sin m\phi\right). 
\label{eqn.multipole}
\end{eqnarray}
and the moments $a_m$ and $b_m$ are
\begin{eqnarray}
a_m(R) & = & \int_{R^{\prime}\le R} \Sigma(\vec x^{\prime})
\left(R^{\prime}\right)^m \cos m\phi^{\prime} d^2x^{\prime}, \nonumber\\
b_m(R) & = & \int_{R^{\prime}\le R} \Sigma(\vec x^{\prime})
\left(R^{\prime}\right)^m \sin m\phi^{\prime} d^2x^{\prime}, \nonumber
\end{eqnarray}
where $\vec x^{\prime} = (R^{\prime},\phi^{\prime})$ and $\Sigma$ is the surface mass density.  In the case of X-ray studies, X-ray surface brightness replaces surface mass density in the calculation of the power ratios.  

The powers are formed by integrating the magnitude of $\Psi_m$, the \textit{m}th term in the multipole expansion of the potential given in equation (12), over a circle of radius $R$,
\begin{equation}
P_m(R)={1 \over 2\pi}\int^{2\pi}_0\Psi_m(R, \phi)\Psi_m(R, \phi)d\phi.
\end{equation}
Ignoring factors of $2G$, this gives
\begin{equation}
P_0=\left[a_0\ln\left(R\right)\right]^2 \nonumber
\end{equation}
\begin{equation}
P_m={1\over 2m^2 R^{2m}}\left( a^2_m + b^2_m\right). \nonumber
\end{equation}
The higher multipole moments are sensitive to the amount of
substructure in the cluster. In most cases, as in this work, we
normalize the multipole moments $P_2$, $P_3$ and $P_4$ to $P_0$ to factor out the
overall brightness of the cluster. $P_2$, the second multipole moment is
sensitive to deviations from circularity, $P_3$ to deviations from mirror
symmetry, and $P_4$ is similar to $P_2$, but sensitive to smaller scale asymmetry. Each of
the moments has a component of radial decline in the cluster profile
which contributes to its value, such that substructure at large radius
leads to larger power ratios.  Perfectly round clusters (irrespective
of the shape of the radial profile) would give all zero power ratios. While each of the power ratios is
sensitive to slightly different asymmetries, it is important to
note that they are strongly correlated with one another. That means
that clusters that are disturbed via merging activity typically show
deviations from circularity, as well as from mirror symmetry and
small scale substructure simultaneously. In this work, we use the
values of the power ratios inside a radius of $r_{500}$, a radius
which is typically observable in X-rays for a large set of galaxy
clusters. We note that the value of the power ratio selected is
somewhat dependent on the outer radius used, and that merging clusters
with disturbances on larger scales than the outer radius may be missed
when using $r_{500}$ \citep{takizawa_NM}. 

A second type of quantitative measure we employ in this study is the
centroid shift \citep[see][]{mohr93,ohara06,poole,maughan}. Centroids
of the X-ray surface brightness are measured successively in apertures
of increasing size. The centroid shift is the variation of the
centroid with radius \citep[see above references and][]{jeltema}, and
is an indicator of deviations from dynamical equilibrium. Our main
interest here is in the change of the mean centroid
shifts from the non-cold front clusters to the cold front clusters.
\begin{table}
\caption{Median Power Ratios and Centroid Shifts \label{structure}}
\begin{tabular}{ccc}
\hline
\hline
Measure & Cold Front Clusters & Non-Cold Front Clusters\\
\hline
$P_2$/$P_0$ & 1.75e-6 & 5.71e-7 \\
$P_3$/$P_0$ & 2.25e-8  & 4.02e-9\\
$P_4$/$P_0$ & 6.50e-9   &8.57e-10\\
Centroid Shift & 0.0146  & 0.0080 \\
\hline
\end{tabular}
\end{table}

We find that clusters identified as having cold fronts have
systematically higher values for the power ratios and centroid shifts
than those that do not have cold fronts.  This result is statistically
significant at $>99$\% when a Kolmogorov-Smirnov test is
applied. Table \ref{structure} shows the median values for the power
ratios and centroid shifts for clusters identified as cold front
clusters and those without cold fronts. Figures \ref{p3} 
and \ref{p4} show the distribution of $P_3$/$P_0$ and centroid shifts
in the cluster
sample for both cold front and non-cold front clusters in the sample
at $z$=0.  The other power ratios have very similar distributions.

\begin{figure}
\begin{center}
\plotone{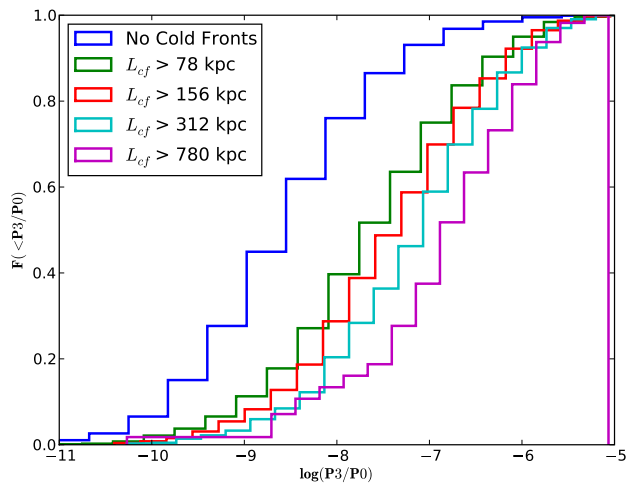}
\end{center}
\caption{Cumulative fraction of cluster images as a function of $P_3$/$P_0$ power
  ratio for clusters with increasing threshold for number of cold
  front pixels identified. Blue line represents all cluster images
  with no cold front pixels identified. Red line represents 5
  identified pixels ($L_{cf}>78h^{-1}$kpc), green line is for 10 pixels
  ($L_{cf}>156h^{-1}$kpc), cyan line is for 20 pixels ($L_{cf}>312h^{-1}$kpc),
  purple line is for images with 50 or more cold front pixels ($L_{cf}
  >780h^{-1}$kpc). Note the shift in the median of the
distribution to more disturbed (higher) $P_3$/$P_0$ values as the threshold
for minimum number of cold front pixels is raised.}
\label{p3steps}
\end{figure}

This result holds for all clusters with even a single pixel identified as a cold
front, but it becomes monotonically more significant when we increase
the threshold for number of pixels identified as cold fronts. This measure is equivalent
to total length of cold front features. Figure \ref{p3steps} shows a
histogram of the $P_3$/$P_0$ power ratio for the clusters in samples with
increasing numbers of cold front pixels. As we raise the threshold for
number of pixels identified as cold fronts, the morphology becomes
systematically more disturbed, as is expected if cold fronts are
related to merging events. This trend is repeated in the other power
ratios, as well as the centroid shift measure. In Figure
\ref{p3steps}, the number of cold front pixels identified in the $z$=0
simulated clusters increases as the distribution shifts to the more disturbed
(right) side of the distribution. As noted in earlier sections, $L_{cf}$ can be
thought of roughly as a total extent of cold front features, where in
these images, the pixel scale is 15.6$h^{-1}$kpc. 

\section{Effects of Baryonic Physics}\label{sec:resolution}
Two other simulations are used
to study the effect of additional non-gravitational physics on the
properties of cold fronts. While not a convergence study, these
results are presented to give a basic indication of the impact of
additional baryonic physics on the result.  We know observationally
that in galaxy clusters the gas physics is not purely adiabatic. It is
well-known that in the core regions of clusters, radiative cooling
becomes important, in many clusters the estimated cooling times are
significantly shorter than a Hubble time. In addition, there are both
heat and metals injected from cluster galaxies due to star formation,
and there is also strong evidence of AGN activity. In this case, we
explore the result of including additional physical processes on the
incidence of cold fronts in the simulated clusters. 
\begin{figure}
\begin{center}
\plotone{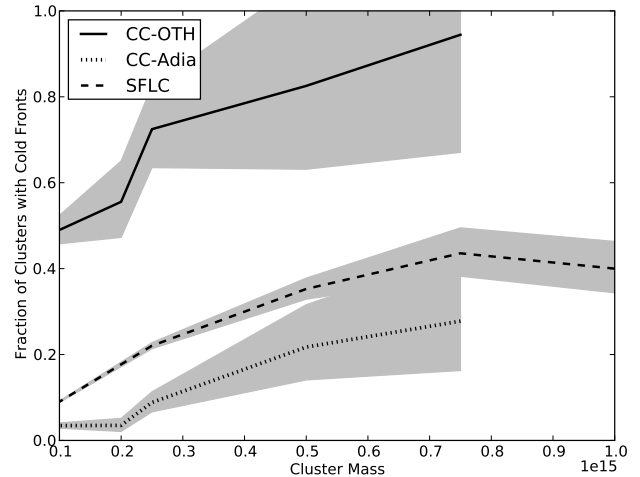}
\end{center}
\caption{Fraction of cluster projections hosting at least 10
pixels ($L_{cf} > 156h^{-1} kpc$) identified by the automated cold front finder. Solid line is for
the CC-OTH run described in the text, with radiative cooling and star
formation plus feedback. Dashed line represents the SFLC simulation, and
is replotted data from Figure \ref{fraction_fm_all}. Dotted line is for
the CC-Adia run, identical to CC-OTH, but with only adiabatic gas
physics. Gray regions are 1$\sigma$ Poisson error bars.}
\label{multi_run}
\end{figure}

\begin{figure*}
\begin{center}
\epsscale{1.0}
\plotone{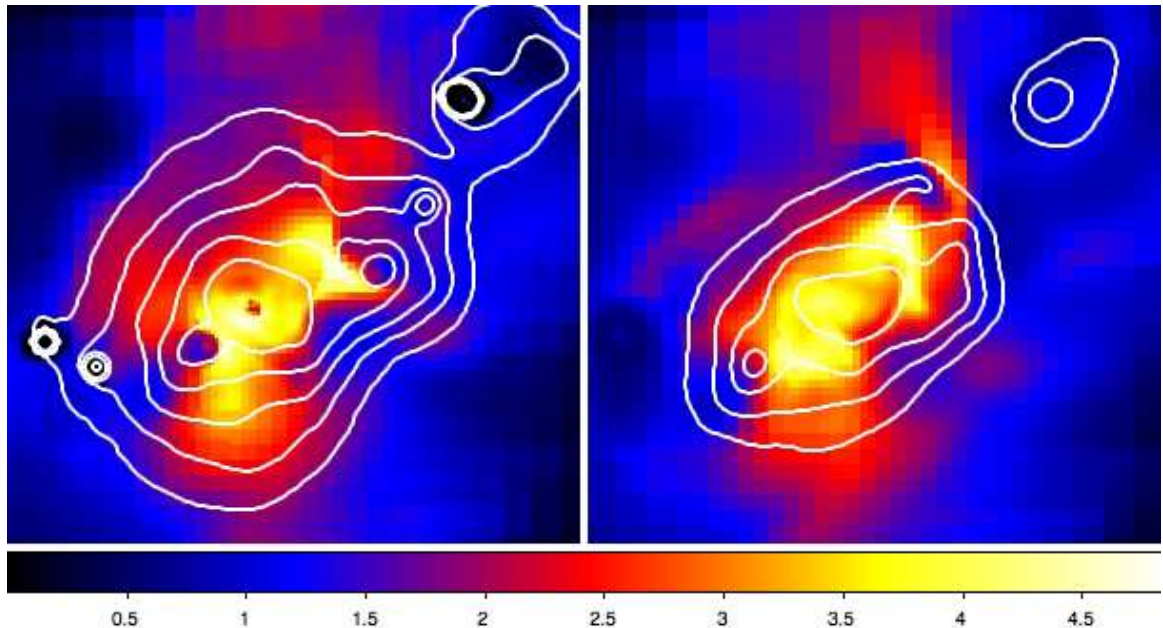}
\end{center}
\caption{Comparison of $T_{sl}$ maps for the same cluster in two
  different simulations, the left panel is from the simulation
  designated CC-OTH and the right is from CC-Adia. $T_{sl}$ is the image
  quantity, the contours are synthetic 0.3-8.0 keV X-ray surface brightness
  contours with the same limits in each map. The 6.1$\times 10^{14}
  M_{\odot}$ cluster from the CC-OTH (simulation with
  radiative cooling, star formation and thermal/metal feedback) in the
  left panel, and the same cluster in the adiabatic simulation is shown 
  in the right panel.  Image scale is roughly 3$h^{-1}$Mpc
  square for each panel. Note the significant difference in the
  temperature maps, with the cold subclumps very obvious in the CC-OTH
  cluster map. Note that the cold subclumps create peaks in
  the X-ray contours.}
\label{image_comp}
\end{figure*}

Our main result is shown in Figure \ref{multi_run}, showing the
fraction of clusters with at least 10 cold front pixels (in this case
the images have the same pixel scale across all simulations) as a
function of cluster mass. The plot axes
match Figure \ref{fraction_fm_all}, and the result from that figure
are overplotted with the same result from the CC-Adia and CC-OTH
simulations. Several things are obvious in this comparison, the first
is that these simulations probe different cluster mass ranges, since
the SFLC represents a much larger physical volume than
CC-Adia and CC-OTH. The second is that including radiative cooling and
feedback results in significantly more clusters hosting cold fronts.
This is not unexpected, as the cooling of cluster cores, as well as
subclusters, results in more cold edges surrounded by hot gas. Figure
\ref{image_comp} illustrates this effect, comparing the identical
cluster in both the CC-Adia and CC-OTH runs. 
\epsscale{1.3}

\begin{figure}
\begin{center}
\plotone{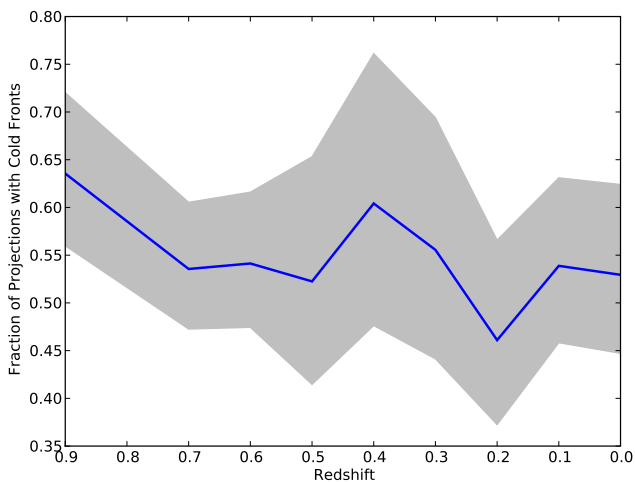}
\end{center}
\caption{Fraction of clusters with M$\geq 10^{14}$ in the CC-OTH simulation with cold front
  length $L_{cf}>156h^{-1} kpc$ as a function of redshift. Gray
  region is 1$\sigma$ Poisson error bars.}
\label{cc_oth_mass}
\end{figure}

Figure \ref{cc_oth_mass} shows the trending of fraction of cold front
clusters with redshift in our simulation with additional baryonic
physics (CC-OTH). While the fraction has some variation with redshift,
within the statistical errors of the sample the trend is roughly
flat.  This compares well with the SFLC adiabatic run, though there is
a mild increase in the fraction of cold front clusters from high to
low redshift in that simulation. In CC-OTH, the fraction goes from
roughly 0.64 at $z$=0.9 to 0.54 at $z$=0.  However, as noted, the Poisson
errors are large ($\sim \pm$0.1 at all $z$) due to the smaller cluster
sample in this simulation.

Additionally we can compare the fraction of clusters with cold fronts
from SFLC to CC-Adia, since the physics is identical, though the mass
resolution is different. CC-Adia has a lower incidence of
cold front clusters than does SFLC, which is somewhat surprising. However, given the
number of changes between SFLC and CC-Adia, it is premature to make
strong conclusions. For example the much reduced box size in our
smaller simulations (from 512$h^{-1}$Mpc to 128$h^{-1}$Mpc cubic
volume) leads to a radical reduction in large-scale power in the
simulation box. What has not changed between any of these runs,
however, is the increasing trend of fraction of cold front clusters with mass.  
\section{Discussion and Summary}\label{sec:discuss}
We have shown the incidence of X-ray cold fronts in a large sample of
simulated galaxy clusters. In this initial study, to
maximize the number of clusters in the sample at useful grid
resolution, we gather the statistical results from an adiabatic
simulation. We identify cold fronts using projected maps
of spectroscopic-like temperature ($T_{sl}$ ), and X-ray
surface brightness generated with Cloudy. We show the effect of additional baryonic physics using
two smaller volume simulations, identical in initial conditions, one
using adiabatic physics, and the other including the effects of
radiative cooling, star formation and metal/thermal feedback. The main
results are summarized here.

\begin{itemize}
\item The trends of simulated galaxy clusters with cold front features
  with redshift depends significantly on cluster selection. When using a mass-limited sample the trend out to $z$=1.0 is weak, irrespective of the threshold of
  number of identified cold front pixels.  This weak trending
  occurs only when the sample is mass-limited, however.  When limiting
  the sample by the rarity of peaks using a constant $\sigma_M$, the
  incidence of cold fronts shows a decreasing trend with redshift, and
  when using a flux limit derived from X-ray scaling relations, the
  trend is increasing with redshift.
\item There is in the simulated cold front clusters a strong trend
  with cluster mass, the more massive clusters being much more likely
  to host a cold front. In the adiabatic sample, the fraction of
  clusters hosting cold fronts with $L_{cf} > 156h^{-1} kpc$ with
  M$>7.5 \times 10^{14} M_{\odot}$ is 40-50\% depending on the
  redshift of the sample. In the simulation with radiative cooling,
  star formation and thermal feedback, that fraction can be higher
  than 80\% at the high mass end. 
\item The fraction of clusters hosting cold fronts in the adiabatic simulations
  is lower than in observed samples, which give fractions anywhere
  from 40-80\% depending on the sample and the redshift coverage. We
  know given our method that we probably underestimate the total
  number of cold fronts, so this is not surprising. In
  our flux-limited sample from the simulations, we find 15\% of
  cluster with $L_{cf} > 156h^{-1}$kpc. We show that including
  radiative cooling and star formation in the simulations will
  increase this fraction significantly without changing the general
  trends with mass and redshift. Additionally as noted above, the
  highest mass clusters in our sample show a higher cold front
  fraction, around 40-50\%. 
\item Cold fronts in simulated clusters are almost universally the
  result of mergers of various sizes, and are almost always associated
  with shocks. This result is still qualitative, but we hope to make
  more quantitative claims in subsequent work.
\item The association of cold front clusters with mergers is confirmed
  in a statistical sense using the quantitative morphological measures
  for the X-ray surface brightness maps. Clusters with cold fronts are
  preferentially more disturbed when measured using power ratios and
  centroid shifts. The significance of this result increases if we
  select clusters with higher numbers of identified cold front pixels.
\item There is a clear impact on the incidence of cold fronts due to
  changes in baryonic physics of the simulation.  When including
  radiative cooling and star formation and feedback to the
  simulations, the incidence of cold fronts goes up strongly to
  greater than 50\% of clusters at all masses. While the \textit{normalization}
  of the number of cold fronts in simulated clusters is strongly
  affected by the baryonic physics, the \textit{trending} with mass
  and redshift is not.  
\end{itemize}

The most obvious improvements that can be made to this analysis are
a full resolution and convergence study with variations in baryonic
physics, and an analysis of the statistics of cold fronts
expected with more realistic X-ray images and temperature maps
including X-ray instrumental effects and backgrounds.

Since it is clear from our analysis of the additional small box
simulations that our results depend on box size and baryonic physics, the first step is most
obvious.  However, the statistics of the cold front clusters are only
useful if they can be related to observational results. The effect of
the length of exposure, redshift of the cluster, angular size of the
cluster in relation to the chip size and spatial response, and the
spectral response all will play a role in detectability of these
features. 
 
In addition, our result regarding the redshift trends of cold front
clusters is clearly a result
of sample selection. What the ``right'' sample selection should be is
unclear, but for observational comparison, it is appropriate to use
flux-limited samples. We believe the result on the incidence as a
function of mass to be robust at this point, since the two additional
simulations show the same trend, though with differing normalization. 
Finally, further study can illuminate how cold fronts can be used as a diagnostic of recent
merger history.  A full study of how the properties of the cold fronts
in any given cluster can be used to deduce recent merger history will
also be the topic of future work.
\acknowledgments
Computations described in this work were performed using the Enzo code
developed by the Laboratory for Computational Astrophysics at the
University of California in San Diego (http://lca.ucsd.edu).  
EJH acknowledges support from NSF AAPF AST-0702923.
EJH, SWS and JOB have been supported in part by a grant 
from the U.S. National Science Foundation (AST-0807215). SWS has been supported
by a DOE Computational Science Graduate Fellowship under grant number DEFG02-
97ER25308.
TEJ has been supported in part by Chandra award AR0-11016B. 
BWO has been supported in part by a grant from the NASA ATFP
program (NNX09AD80G). BWO has been funded in part
under the auspices of the U.S.\ Dept.\ of Energy, and supported by its
contract W-7405-ENG-36 to Los Alamos National Laboratory.  Some
simulations were by performed at SDSC and NCSA with computing time provided by 
NRAC allocation MCA98N020. CC-Adia and CC-OTH simulations and
post-processing analysis were performed at
NICS on Kraken, and at TACC on Ranger using NSF TeraGrid allocations
TG-AST090040 and TG-AST100004. EJH thanks
Marcus Bruggen, Evan Scannapieco, Maxim Markevitch, John ZuHone, and
Ryan Johnson for useful discussions. The authors thank the anonymous
referee for their very useful comments.

\bibliographystyle{apj}

\end{document}